\documentclass[twocolumn,showpacs,pra,floatfix]{revtex4}

\usepackage{graphicx}
\usepackage{bm}
\usepackage{psfrag}
\usepackage{amsmath}
\usepackage{amssymb}
\usepackage{latexsym}
\usepackage{exscale}
\usepackage{pdfsync}
\newcommand{\vect}[1]{\bm{#1}}
\newcommand{\ten}[1]{\mbox{\textbf{{\textsf{#1}}}}}

\newcommand{\trace}{\operatorname{tr}}

\newcommand{\dif}{\mathrm{d}}
\newcommand{\re}{\mathrm{Re}}

\newcommand{\mi}{\mathrm{i}}
\newcommand{\me}{\mathrm{e}}
\newcommand{\rr}{\vect{r}}
\newcommand{\ra}{\vect{r}_{\!A}}
\newcommand{\rb}{\vect{r}_{\!B}}
\begin{document}

\title{Body-assisted van der Waals interaction between
excited atoms}

\author{Hassan Safari}
\affiliation{Graduate University of Advanced Technology, Ending of  
Haft-Bagh Highway, Kerman, Iran}
\author{Mohammad Reza Karimpour}
\affiliation{Graduate University of Advanced Technology, Ending of 
Haft-Bagh Highway, Kerman, Iran}

\date{\today}
\begin{abstract}
We present a formula for the body-assisted van der Waals interaction 
potential between two atoms, one or both being prepared in 
an excited energy eigenstate. The presence of arbitrary arrangement 
for material environment is taken into account via the Green 
function. The resulting formula supports one of two conflicting 
findings recorded. The consistency of our formula is investigated by 
applying it for the case of two atoms in free space and comparing the 
resulting expression with the one found from the limiting 
Casimir-Polder potential between an excited atom and a small 
dielectric sphere. 
 
\end{abstract}

           
\maketitle
%%%%%%%%%%%%%%%%%%%%%%%%%%%%%%%%%%%%%%%%%%%%%%%%%%%%%%%%%%%%%%%%%%%%%
Dispersion interactions are understood as a result of quantum 
description of the electromagnetic field. As the atoms under 
consideration are assumed to be ground-state atoms, transitions 
to excited energy eigenstates are involved in the interaction 
together with the emission of virtual photons from a continuous range
of frequencies. In the case of excited atoms, transitions to 
lower-lying states can occur while releasing the energy difference in 
the form of real photons of certain discrete frequencies (see, e.g., 
Ref.~\cite{Wylie85} for atom-body interaction). In the two-atom case,
the long-range potential was first found to have an oscillatory 
distance-dependence with an amplitude falling off as $r^{-2}$ 
\cite{Gomberoff66,McLone65}. Later on, the long-range potential was 
confirmed to fall of as $r^{-2}$, but without oscillations 
\cite{Power93,Power95}. The difference between the two findings arises 
from the way the photon integrals have been treated. The two 
contradictory results are valid from mathematical point of view and a 
possible oscillatory behaviour of the retarded van der Waals (vdW) 
potential remains as an open question. A time-dependent calculation 
supports the oscillatory result~\cite{Rizzuto04}.
On the experimental side, the vacuum-induced level shift (Casimir-Polder potential)
of an excited barium ion in the presence of a mirror,
is observed to show an oscillatory distance-dependence \cite{Wilson, Bushev}.

In this letter, we first derive the body-assisted vdW interaction 
potential between two excited atoms, using fourth-order perturbation 
theory. As the background media is 
replaced by free space in our formula, the resulting expression for 
the long-range interatomic separation shows an oscillatory 
distance-dependence, in agreement with
Refs.~\cite{Gomberoff66,McLone65}, while 
it does not agree with the formula given in 
Refs.~\cite{Power93, Power95}. 
In order to facilitate a judgement about the two contradictory 
results, we will take a proper limit from the known atom-body 
Casimir-Polder (CP) potential for an excited atom and reduce it to the 
atom-atom vdW interaction, to see whether the outcome supports any of 
the above mentioned results. 

Lets first, derive the formula of the body-assisted vdW interaction 
between two excited atoms. A very detailed derivation of the formula 
for the case of ground-state atoms is given in Ref.~\cite{HS06}. In 
the case of excited atoms, the starting point 
is the same as for the ground-state atoms, hence we refer the reader 
to the calculation in Ref.~\cite{HS06}. We only point the differences 
which arise from the fact that one or both atoms might be in excited 
states here.

Consider two atoms $A$ and $B$ in the presence of an arbitrary 
arrangement of magneto-electric bodies, located at positions $\rr_{A}$ 
and $\rr_{B}$, each being excited to an energy eigenstate, say 
$|k\rangle_{\!A}$ and  $|l\rangle_{\!B}$, respectively. The vdW 
interaction potential resulting form the fourth-order perturbation, 
following a calculation similar to the one for the ground-state 
atoms in Ref.~\cite{HS06}, leads to
\begin{multline}
\label{eq35}
U_{AB}(\ra,\rb)
  =\frac{\mi\mu_0^2}{\hbar\pi}
         \sum_{
 \genfrac{}{}{0pt}{}{m\neq k}{n \neq l} }
  \frac{1}{\omega_A^{mk}+\omega_B^{nl}}\\
 \hspace{-.4in}\times \mathcal{P} \biggl\{
     \int_0^\infty \dif\omega
     \frac{\omega^4 (\omega_A^{mk}+\omega_B^{nl}
     +\omega)}
     {(\omega+\omega_A^{mk}) (\omega+\omega_B^{nl})}\\
     \hspace{.3in}+ \int_0^{-\infty} \dif \omega
     \frac{\omega^4 (\omega_A^{mk}+\omega_B^{nl}-\omega)}
     {(\omega-\omega_A^{mk}) (\omega-\omega_B^{nl})}\biggr\}
    \\
    \times [\vect{d}_A^{km}\cdot
     \ten{G}(\rb,\rb,\omega)
      \cdot \vect{d}_B^{ln}]^2
\end{multline}
(compare with Eq.~(45) in Ref.~\cite{HS06}) with $
{\vect{d}}_{\!A}^{km}
 = \langle k|
\hat{\vect{d}}_{A}
 |m\rangle$ and $\omega_{\!A}^{km}$ denoting, respectively, the 
 electric dipole moments and frequencies of the atomic transitions. 
 All geometric and magneto-electric 
 properties of the environmental media are contained 
  in the  Green tensor $\ten{G}$ via the 
 frequency-dependent relative electric permittivity 
 $\varepsilon(\rr,\omega)$ and relative 
 magnetic permeability $\mu(\rr,\omega)$. The Green tensor is the 
 unique solution to the 
 inhomogeneous Helmholtz differential equation
\begin{align}
\nabla\times &\frac{1}{\mu(\rr,\omega)}\nabla\times
\ten{G}(\rr, \rr',\omega)
-\varepsilon(\rr,\omega)\frac{\omega^2}{c^2}\ten{G}(\rr, \rr',
\omega)\nonumber\\
&=\ten{I}\delta(\rr-\rr')
\end{align}
with the boundary condition
\begin{equation}
\ten{G}(\rr, \rr',\omega)=\ten{0}\quad \mathrm{for}\, |\rr-\rr'|
\to\infty\,.
\end{equation}
Further, it obeys the Schwartz 
reflection principle,
\begin{equation}
\label{Schw}
\ten{G}(\rr,\rr',\omega) = \ten{G}^\ast(\rr,\rr',-\omega^\ast),
\end{equation}
and Onsager reciprocal relation,
\begin{equation}
\label{Ons}
\ten{G}(\rr,\rr',\omega) = \ten{G}^\top(\rr',\rr,\omega).
\end{equation}

The integrands in Eq.~(\ref{eq35}), recalling 
the general properties the Green tensor $\ten{G}$ as a response 
function, are analytic for $m>k$ and 
$n>l$, in the upper half of the complex-frequency plane including the 
real axis. The letter $\mathcal{P}$ just before the curly brackets 
stands 
for principal value and makes a particular sense for $m<k$ and/or 
$n<l$. Equation~(\ref{eq35}) can be simplified by using 
contour-integral techniques. For the first integral, we may use 
Cauchy's theorem and replace the 
integral by a contour integral
along infinitesimal half-circles around the possible poles at $\omega 
= -\omega_A^{mk}= \omega_A^{km}$ and $\omega 
= -\omega_B^{nl}= \omega_B^{ln}$, an 
infinitely large quarter-circle in the first quadrant and along the
positive imaginary axis, introducing a purely imaginary frequency,
$\omega$ $\!=$ $\!\mi u$. The integral along the infinitely large
quarter-circle vanishes due to the limiting behaviour of the Green 
tensor \cite{Knoll2001} 
\begin{equation}
\label{eq36}
\lim_{|\omega| \rightarrow \infty}
\frac{\omega^2}{c^2} \ten{G}(\rr,\rr',
\omega)=-\ten{I}\delta(\rr-\rr').
\end{equation}
The result becomes
\begin{eqnarray}
\label{eq37}
&&\mathcal{P}\int_0^\infty \dif\omega
     \frac{\omega^4 (\omega_A^{mk}+\omega_B^{nl}
     +\omega)}
     {(\omega+\omega_A^{mk}) (\omega+\omega_B^{nl})}\mathcal{G}
     ({\rr_{A}},{\rr_{B}},
     \omega)\nonumber\\
&& =\int_0^\infty \dif u  
     \frac{\mi u^4 (\omega_A^{mk}+\omega_B^{nl}
     +\mi u)}
     {(\mi u+\omega_A^{mk}) (\mi u+\omega_B^{nl})}
    \mathcal{G}({\rr_{A}},{\rr_{B}},
     \mi u)\nonumber\\
&&+\mi \pi \frac{\omega_A^{km}\omega_B^{ln}}
{\omega_B^{ln}-\omega_A^{km}}
\left[\Theta(k-m)\left(\omega_A^{km}\right)^3
\mathcal{G}(\rr_{A},\rr_{B}, \omega_A^{km})\right.
     \nonumber\\
&&%\hspace{.5in}
- \left.\Theta(l-n)\left(\omega_B^{ln}\right)^3
\mathcal{G}({\rr_{A}},{\rr_{B}},
 \omega_B^{ln})\right]%\nonumber\\
\end{eqnarray}
[$\Theta(x)$, unit step function], where  
$\mathcal{G}(\rr_{A},\rr_{B},
 \omega)$ is used as an abbreviation for 
 $[\vect{d}_A^{km}\cdot\ten{G}(\rr_{A},\rr_{B},\omega)
 \cdot \vect{d}_B^{ln}]^2$.
 In a similar manner, for the second integral in Eq.~(\ref{eq35}) we 
  find
\begin{eqnarray}
\label{eq38}
&&\mathcal{P}\int_0^{-\infty} \dif\omega
     \frac{\omega^4 (\omega_A^{mk}+\omega_B^{nl}
     -\omega)}
     {(\omega-\omega_A^{mk}) (\omega-\omega_B^{nl})}\mathcal{G}
     (\rr_{A},\rr_{B}, \omega)\nonumber\\
&& =\int_0^\infty \dif u  
     \frac{\mi u^4 (\omega_A^{mk}+\omega_B^{nl}
     -\mi u)}
     {(\mi u-\omega_A^{mk}) (\mi u-\omega_B^{nl})}
    \mathcal{G}(\rr_{A},\rr_{B},\mi u)\nonumber\\
&&+\mi \pi \frac{\omega_A^{km}\omega_B^{ln}}
{\omega_B^{ln}-\omega_A^{km}}
\left[\Theta(k-m)\left(\omega_A^{km}\right)^3
\mathcal{G}^\ast(\rr_{A},\rr_{B},\omega_A^{km})\right.
     \nonumber\\
&&%\hspace{.5in}
- \left.\Theta(l-n)\left(\omega_B^{ln}\right)^3
\mathcal{G}^\ast(\rr_{A},\rr_{B},
 \omega_B^{ln})\right],%\nonumber\\
\end{eqnarray}
where Eq.~(\ref{Schw}) is used. Now, by combining Eqs.~(\ref{eq35}), 
(\ref{eq37}) and (\ref{eq38}), and making use of  
Eq.~(\ref{Ons}), the two-atom interaction potential, 
after being split into the off-resonant and resonant parts, can be 
written as follows:  
\begin{equation}
U(\ra,\rb) = U^\mathrm{or}(\ra,\rb)+U^\mathrm{r}(\ra,\rb),
\end{equation}
\begin{eqnarray}
\label{eq40}
&&U^\mathrm{or}(\ra,\rb)=-\frac{\hbar\mu_0^2}{2\pi}
\int_0^\infty \dif u \,u^4\nonumber\\
&&\times\trace\left[\bm{\alpha}_A^k(\mi u)
\cdot
\ten{G}(\ra,\rb,\mi u)\cdot \bm{\alpha}_B^l(\mi u)\cdot
\ten{G}(\rb,\ra,\mi u)
\right],\nonumber\\
\end{eqnarray}
\begin{eqnarray}
\label{eq41}
&&\hspace{-.1in}U^\mathrm{r}(\ra,\rb) \nonumber\\
&&\hspace{.1in}= 
-\mu_0^2\sum_{m<k}\left(\omega_A^{km}\right)^4\re\,
\left[\vect{d}_A^{km}
\cdot\ten{G}(\ra,\rb,\omega_A^{km})\right.\nonumber\\
&&\hspace{.8in}\left.\cdot\bm{\alpha}_B^l(\omega_A^{km})\cdot
\ten{G}(\rb,\ra,\omega_A^{km})\cdot\vect{d}_A^{mk}\right]\nonumber\\
&&\hspace{.3in}-\mu_0^2\sum_{n<l}\left(\omega_B^{ln}\right)^4
\re\left[\vect{d}_B^{ln}
\cdot\ten{G}(\rb,\ra,\omega_B^{ln})\right.\nonumber\\
&&\hspace{.8in}\left.\cdot\bm{\alpha}_A^k(\omega_B^{ln})\cdot
\ten{G}(\ra,\rb,\omega_B^{ln})\cdot\vect{d}_B^{nl}\right].
\end{eqnarray}
In Eqs.~(10) and (11), $\bm{\alpha}_{\!A}^{k}(\omega)$ is the electric 
polarizability tensor of atom $A$ in the $k$-th energy eigenstate, 
defined as
\begin{align}
\label{eq4}
\bm{\alpha}_{\!A}^{k}(\omega)&= \frac{2}{\hbar}\lim_{\epsilon\to 0+}
    \sum_m
  \frac{\omega_{\!A}^{mk} \vect{d}_{\!A}^{km}\vect{d}_{\!A}^{mk}}
      {(\omega_{\!A}^{mk})^2-\omega^2-\mi\epsilon\omega}\nonumber\\
     & = \frac{2\ten{I}}{3\hbar}\lim_{\epsilon\to 0+}
    \sum_m
  \frac{\omega_{\!A}^{mk} |\vect{d}_{\!A}^{km}|^2}
      {(\omega_{\!A}^{mk})^2-\omega^2-\mi\epsilon\omega}\equiv 
      \alpha_{\!A}^k(\omega) \ten{I}\,.
\end{align}

Needless to say that in the case where atom $B$ is in its 
ground state ($l=0$), 
the second term on the right-hand side of Eq.~(\ref{eq41}) vanishes. 

In the simplest case of two isotropic atoms in an infinitely extended 
free space, the required Green tensor ($\ten{G}\to\ten{G}^{(0)}$) is 
given as
\cite{Knoll92}
\begin{multline}
\label{eq310}
\ten{G}^{(0)}(\rr_{\!A},\rr_{\!B},\omega) \\= \frac{-c^2\me^{\mi 
\omega 
l/c}}{4\pi \omega^2 l^3} \left[p(-\mi l\omega/c)\ten{I} -q(-\mi 
l\omega/c)\mathbf{e}\mathbf{e}\right]
\end{multline}
with $l=|\rr_B-\rr_A|$, $\mathbf{e} = (\rr_B-\rr_A)/l$ and
\begin{align}
\label{eq320}
p(x)& = 1+x+x^2,\\
\label{eq33}
q(x)& = 3+3x+x^2.
\end{align}

Choosing the Cartesian coordinates system 
such that its origin corresponds to the location of atom $A$ while 
$\rr_B = (r,0,0)$, the only non-zero matrix elements of the 
Green tensor are as follows:
\begin{equation}
\label{eq31}
\hspace{-.6in}G_{xx}^{(0)}(\rr_{\!A},\rr_{\!B},\omega) = \frac{c^2}
{2\pi \omega^2 r^3} (1-\mi r\omega/c)\me^{\mi \omega r/c},
\end{equation}
\begin{multline}
\label{eq32}
G_{yy}^{(0)}(\rr_{\!A},\rr_{\!B},\omega) = G_{zz}^{(0)}(\rr_{\!A},
\rr_{\!B},\omega)\\
 =  \frac{-c^2}{4\pi \omega^2 r^3} (1-\mi r\omega/c-
 r^2\omega^2/c^2)\me^{\mi 
\omega r/c}.
\end{multline}
Substitution of these into Eqs.~(\ref{eq40}) and (\ref{eq41})   
leads to
\begin{equation}
\label{eq23}
U^{\mathrm{or}}(r) = 
 \frac{-\hbar}{16\pi^3\varepsilon_0^2r^6}\int_0^\infty\dif u\, 
 \alpha^k_{\!A}(\mi u)\alpha_{B}^0(\mi u)f(ru/c),
\end{equation}%
\begin{eqnarray}
\label{eq27}
&&\hspace{-.1in}U^{\mathrm{r}}(r)  = \frac{-1}{24\pi^2
\varepsilon_0^2 r^6}
\sum_{m<k}|\vect{d}_{\!A}^{km}|^2\alpha_{B}^0
(\omega_{\!A}^{km})\nonumber\\
&&\times 
\left[(3\!-\!5\eta_m^2\!+\!\eta_m^4)\cos(2\eta_m)
+(6\eta_m\!-\!2\eta_m^3)\sin(2\eta_m)\right],\nonumber\\
\end{eqnarray}
where $\eta_m=r\omega_{\!A}^{km}/c$, atom $B$ is assumed to be in its 
ground state and
\begin{equation}
f(x)=e^{-2x}(3+6x+5x^2+2x^3+x^4).
\end{equation}
Equation~(\ref{eq23}) is exactly the well known result for the
off-resonant part of the vdW interaction potential in free space 
(see, e.g., Ref.~\cite{Power95}). The resonant part, Eq.~(\ref{eq27}), 
is in agreement with Ref.~\cite{Gomberoff66} in the retarded limit, 
while it does not agree with the finding of 
Refs.~\cite{Power93,Power95}. 
The difference arises from the ways the perturbative calculation was 
accomplished; the frequency integrals in Refs.~\cite{Power93,Power95} 
were performed such that 
the poles on the real axis were treated by addition or subtraction of 
small pure-imaginary frequencies whereas in Ref.~\cite{Gomberoff66} the 
integrals were treated as principal 
value integrals.
 If one imitates the calculations in 
Ref.~\cite{Power95} in order to generalize its result to 
the inclusion of material background, for the resonant part of the vdW 
interaction potential between an excited atom $A$ and a ground-state  
atom $B$, ends up with
\begin{multline}
\label{eq30}
U^{\mathrm{r}}(\rr_{\!A},\rr_{\!B}) 
=-\frac{\mu_0^2}{3}\sum_{m<k}(\omega_{\!A}^{km})^4\\
\times\left|\vect{d}_{\!A}^{km}\right|^2
\alpha_{\!B}^0(\omega_{\!A}^{km}) \sum_{i,j}
\left|G_{ij}(\rr_{\!B},\rr_{\!A},\omega_{\!A}^{km})\right|^2
\end{multline} 
(for an equivalent formula for two-level atoms see Eq.~(68) in 
Ref.~\cite{Sherkunov07}  or Eq.~(3) in Ref.~\cite{Tomas08}). Now, the 
interaction potential in free space, according to this 
formula, can 
be obtained using the matrix elements (\ref{eq31}) and  
(\ref{eq32}). The result, as expected, meets the one given in 
Refs.~\cite{Power93, Power95},
\begin{multline}
\label{eq29}
U^r(r) = \frac{-\mu_0^2}{24\pi^2 r^2}
\sum_{m<k}(\omega_A^{km})^4\\
\times| \vect{d}_{\!A}^{km}|^2\alpha_{\!B}^0
(\omega_{\!A}^{km})\left(1+\eta_m^{-2}+
3\eta_m^{-4}\right).
\end{multline}

A possible way to judge about the two contradictory results, 
Eq.~(\ref{eq27}) and Eq.~(\ref{eq29}), may be attacking the problem 
via a sufficiently different approach. To this end, 
we start with the known
CP interaction formula of an excited atom with a macroscopic body. The 
atom-atom interaction can be found by taking a proper limit of a small 
dielectric body (a homogeneous sphere here) and replacing it with a 
second atom.    
 
According to the findings of Refs.~\cite{Wylie85, Stefan04}, the 
CP 
Potential of an isotropic atom $A$ prepared in an energy eigenstate 
$|k\rangle$ and located at a position $\ra$ is given as
\begin{equation}
\label{CP}
U(\ra)=U^{\mathrm{or}}(\ra) + U^{\mathrm{r}}(\ra),
\end{equation}
where $U^{\mathrm{or}}$ and $U^{\mathrm{r}}$ are, respectively, the 
off-resonant and resonant parts of the potential
\begin{eqnarray}
\label{eq2}
&&\hspace{-3ex}U^{\mathrm{or}}(\ra)=\frac{\hbar\mu_0}{2\pi}
\int_0^\infty\dif u\, u^2\,
\alpha_{\!A}^k(\mi u)
\trace\ten{G}^{(1)}(\ra,\ra,\mi u),\\
\label{eq3}
&&\hspace{-3ex}U^{\mathrm{r}}(\ra)=
 -\frac{\mu_0}{3}\sum_{m<k}\left(\omega_{\!A}^{km}|
\vect{d}_{\!A}^{km}|\right)^2
\trace [\mathrm{Re}\,\ten{G}^{(1)}(\ra,\ra,\omega_{\!A}^{km})]
\nonumber\\
\end{eqnarray}
with $\ten{G}^{(1)}$ being the scattering part of the Green tensor.

 Let's consider the atom at a distance 
$r$ from the center of a 
homogeneous dielectric sphere of radius $a$ ($r>a$). The equi-postion 
Green tensor required in Eqs.~(\ref{eq2}) and (\ref{eq3}) can be 
extracted from  Ref.~\cite{Li94} given for a more general case of 
a spherical magneto-electric multilayer. 
However, we adopt a simplified version from 
Ref.~\cite{HS08}. Choosing the spherical coordinates
system such that its origin coincides with the center of the
sphere, the scattering part of the equi-position Green
tensor reads
\begin{equation}
\label{eq7}
{\ten{G}}^{(1)}(\rr,\rr,\omega)=
\sum_{i=r,\theta,\phi}
 G_{ii}^{(1)}(\rr,\rr,\omega)
 \vect{e}_{i}\vect{e}_{i}
\end{equation}
with $\vect{e}_r$, $\vect{e}_\theta$, and $\vect{e}_\phi$ being the 
 unit vectors
pointing the directions of radial distance $r$, polar angle $\theta$,
and azimuthal angle $\phi$, respectively. The matrix elements 
of the Green tensor (\ref{eq7}) are as follows:
\begin{equation}
\label{eq8}
G^{(1)}_{rr}=\frac{\mi c}{4\pi \omega r^2}\sum_{n=1}^\infty
n(n\!+\!1)(2n\!+\!1)B_n^N(\omega)\left[h_n^{(1)}(r\omega/c)\right]^2,
\end{equation}
\begin{multline}
\label{eq9}
G^{(1)}_{\theta\theta} =G^{(1)}_{\phi\phi} =\frac{\mi\omega}{8\pi  c}
 \sum_{n=1}^\infty (2n\!+\!1)\left\{B_n^M(\omega)\left[h_n^{(1)}
 (r\omega/c)\right]^2\right.\\
\left.+\frac{c^2B_n^N(\omega)}
 {\omega^2 r^2}
\left[z h_n^{(1)}(z)\right]_{z=r\omega/c}^{\prime\,2}\right\},
\end{multline}
where
$h_n^{(1)}(z)$ denotes the spherical Hankel function of the first 
kind, prime indicates differentiation with respect to the argument, 
and
\begin{equation}
\label{BM}
B_n^M(\omega)=-\frac{[z_0j_n(z_0)]^\prime
  j_n(z_1)\!-\![z_1j_n(z_1)]^\prime
  j_n(z_0)}{[z_0h^{(1)}_n(z_0)]^\prime
  j_n(z_1)\!-\![z_1j_n(z_1)]^\prime
  h^{(1)}_n(z_0)}\,,
\end{equation}
\begin{equation}
\label{BN}
B_n^N(\omega)=-\frac{\varepsilon(\omega)[z_0j_n(z_0)]^\prime
  j_n(z_1)-[z_1j_n(z_1)]^\prime j_n(z_0)}{\varepsilon(\omega)
  [z_0h^{(1)}_n(z_0)]^\prime j_n(z_1)-[z_1j_n(z_1)]^\prime
  h^{(1)}_n(z_0)}
\end{equation}
with $j_n(z)$ being the spherical Bessel function of the first kind,
$z_0$ $\!=$ $\!a\omega/c$,
$z_1$ $\!=$ $\!\sqrt{\varepsilon(\omega)}z_0$.
Substitution of $\ten{G}^{(1)}$ from Eq.~(\ref{eq7}) together with 
(\ref{eq8}) and (\ref{eq9}) into Eqs.~(\ref{eq2}) and (\ref{eq3}) 
leads to
\begin{align}
\label{eq12}
&U^{\mathrm{or}}(\ra) = \frac{\hbar\mu_0 c}{8\pi^2 r^2}
\sum_{n=1}^\infty(2n\!+\!1)
\int_0^\infty\dif u \,u\, \alpha_{\!A}^k(\mi u)\nonumber\\
&\quad\times\left(B_n^N(\mi u)\left\{n(n\!+\!1)\left[h_n^{(1)}
(z)\right]^2 +\left[zh_n^{(1)}(z)\right]'^{\,2}\right\}\right.
\nonumber\\
&\left.\quad\qquad-\frac{r_{\!A}^2 u^2}{c^2}B_n^M 
(\mi u)\left[h_n^{(1)}
(z)\right]^2\right)_{z=\mi r u/c}
,
\end{align}
\begin{align}
\label{eq13}
U^{\mathrm{r}}&(\ra) = \frac{\mu_0 c}{12 \pi r^2}
\sum_{m<k}\omega_{\!A}^{km}|\vect{d}_{\!A}^{km}|^2
\sum_{n=1}^\infty(2n\!+\!1)
\nonumber\\
&\times\mathrm{Im}\left(
B_n^N(\omega_{\!A}^{km})\left\{
 n(n\!+\!1)\left[h_n^{(1)}(z)\right]^2\!+\!\left[zh_n^{(1)}
 (z)\right]^{\prime\, 2}\right\}\right.\nonumber\\
&\qquad\quad \left. +
\frac{r^2(\omega_{A}^{km})^2}{c^2}
B_n^M(\omega_{\!A}^{km})\left[h_n^{(1)}
(z)\right]^2\right)_{z=r\omega_{\!A}^{km}/c}.
\end{align}
Equations (\ref{eq12}) and (\ref{eq13}) are valid as long as the atom 
is far enough from the surface of the sphere 
such that the description of the molecular structure of the sphere in 
a macroscopic manner is valid. However, we 
are interested in the limiting case of a small sphere, $a\ll r$. 
Following a discussion similar to the one 
given in Ref.~\cite{Stefan03}, it can be shown that, in the 
summands in Eqs.~(\ref{eq12}) and (\ref{eq13}), it is 
enough to retain only the $n=1$ terms  for which, Eqs.~(\ref{BM}) and 
(\ref{BN}) reduce to
\begin{equation}
\label{BM1}
B_1^M(\omega) \simeq 0,
\end{equation}
\begin{equation}
\label{BN1}
B_1^N(\omega) = \frac{2\mi}{3}\frac{\varepsilon(\omega)-1}
{\varepsilon(\omega)+2}\left(\frac{a\omega}{c}\right)^3.
\end{equation}
Substitution of these results for $B_1^M$ and $B_1^N$ and the explicit 
forms of the spherical Hankel function $h_1^{(1)}(z)$ into 
Eq.~(\ref{eq12}) and 
(\ref{eq13}) leads, after some simplifications, to
\begin{equation}
\label{eq16}
U^{\mathrm{or}}(\ra) = \frac{-\hbar a^3}{4\pi^2 \varepsilon_0 
r^6}\int_0^\infty \dif u\,\alpha_{\!A}^k
(\mi u)\frac{\varepsilon(\mi u)-1}{\varepsilon(\mi 
u)+2}f(ru/c),
\end{equation}
\begin{eqnarray}
\label{eq19}
\label{ur}
&&\hspace{-7ex}U^{\mathrm{r}}(\ra) = \frac{-a^3}{6\pi\varepsilon_0 
r^6}
\sum_{m<k} \left|\vect{d}_{\!A}^{km}
\right|^2\frac{\varepsilon(\omega_{\!A}^{km})\!-\!1}
{\varepsilon(\omega_{\!A}^{km})\!+\!2}
\nonumber\\
&&\hspace{-6ex}\times\!\left[(3\!-\!5\eta_m^2\!+\!
\eta_m^4)\cos(2\eta_m)+(6\eta_m\!-\!2\eta_m^3)
\sin(2\eta_m)\right]
\end{eqnarray}

Now, let us consider a sphere to
which the Clausius-Mossotti relation applies, so that
\begin{equation}
\label{eq21}
\frac{\varepsilon(\omega)-1}{\varepsilon(\omega)+2} =  
\frac{\alpha_{\mathrm{s}}(\omega)}{4 \pi a^3\varepsilon_0}
\end{equation}
with $\alpha_{\mathrm{s}}$ being the electric 
polarizability of the sphere.
Making use of Eq.~(\ref{eq21}) in Eq.~(\ref{eq16}) leads back to
 formula (\ref{eq23}), being the
off-resonant part of the vdW interaction potential between an excited
atom $A$ and a ground-state atom in free space, where the electric 
polarizability of the sphere is just replaced by the ground-state polarizability of atom $B$. 

Recovering the off-resonant part of the two-atom interaction potential
via the limiting procedure mentioned above, makes adequately reliable
obtaining also the resonant part via the same approach. This can be 
done by making use of Eq.~(\ref{eq21}) in 
Eq.~(\ref{eq19}). The result coincides with 
Eq.~(\ref{eq27}), which was based on a perturbative calculation, as 
the electric polarizability of the sphere is replaced by that of a 
ground-state atom $B$. Hence, it is evident that the photon integrals must be treated as 
principal-value integrals as in Refs.~\cite{Gomberoff66,McLone65} and 
our formula, 
Eq.~(\ref{eq41}), is the correct one for the resonant vdW interaction 
potential between excited atoms. 

In summary, in this letter we presented a new formula for the 
medium-assisted vdW interaction potential between two excited atoms.
It can be a generalization of the findings of Refs.~\cite{Gomberoff66, 
McLone65} to the presence of arbitrary material 
environment. 
The consistency of the formula was confirmed by comparing its result 
for the case of two atoms in free space with the result 
obtained via a limiting approach from atom-body CP potential. The 
free-space result is of oscillatory distance-dependence behaviour, in 
agreement with Refs.~\cite{Gomberoff66,McLone65} for retarded
atom-atom separations.

\end{document}